\documentclass[a4paper,11pt]{article}
\pdfoutput=1 % if you are submitting a pdflatex (i.e. if you have
\usepackage{xcolor}
\usepackage{jinstpub} % for details on the use of the package, please
\usepackage[normalem]{ulem}
\usepackage[separate-uncertainty,retain-explicit-plus, per-mode = symbol, detect-weight=true, range-units=brackets]{siunitx}
\usepackage{rotating}

\title{Dark Matter Detection Capabilities of a Large Multipurpose Liquid Argon Time Projection Chamber}

\author{E. Church,} %\note{Corresponding author.}}
\author{C.M. Jackson,}
\author{R. Saldanha}

\affiliation{Pacific Northwest National Laboratory, Richland, WA 99354}

\emailAdd{eric.church@pnnl.gov, christopher.jackson@pnnl.gov, richard.saldanha@pnnl.gov}

\newcommand{\artn}{$^{39}$Ar}
\newcommand{\ur}{$^{238}$U}
\newcommand{\thr}{$^{232}$Th}
\newcommand{\alphan}{$(\alpha,$ n)}
\newcommand{\uar}{UAr}
\newcommand{\fnt}{F90}
\newcommand{\psd}{PSD}
\DeclareSIUnit\kgar{kg_\text{Ar}}

\abstract{Liquid Argon Time Projection Chambers are planned to comprise a central role in the future of the U.S. High Energy Physics neutrino program. In particular, this detector technology will form the basis for the 40~kton Deep Underground Neutrino Experiment (DUNE). In this paper we take as a starting point the dual phase far detector design proposed by the DUNE experiment and ask what changes are necessary to allow one of the four 10~kt modules to be sensitive to heavy Weakly Interacting Massive Particle (WIMP) dark matter. We show that with control over backgrounds and the use of low radioactivity argon, which may be commercially available on that timescale, along with a significant increase in light detection, one DUNE-like  module gives a competitive WIMP detection sensitivity, particularly above a dark matter mass of 100 GeV.}

\begin{document}
\setcounter{tocdepth}{2}
\maketitle
\flushbottom

\section{Introduction}
\label{sec:intro}

Liquid Argon Time Projection Chambers (LArTPCs) have been used in neutrino physics in the ICARUS~\cite{ICARUS} experiment and most recently in the MicroBooNE detector~\cite{MicroB}. The culmination of the U.S. LArTPC program will be the construction of the 4-module 40 kt Deep Underground Neutrino Experiment (DUNE) detector, to be located at the 4850 foot level of the Sanford Underground Research Facility near Lead, South Dakota. Data taking for the first two 10 kt modules -- one single liquid phase (SP), and the other dual phase (DP) -- is set to begin in 2024, though the full power neutrino beam is not scheduled to arrive until 2026. The third module will be either SP or DP and the exact technology for the fourth module is currently unspecified. This fourth 10 kt detector is known, therefore, as the module of opportunity \cite{dune_mood}. 

The main charge for investigators putting forward ideas for the module of opportunity is that the primary neutrino physics program of detection of the Charge Parity (CP) phase and the mass hierarchy determination are not compromised.  There are many suggested ideas currently under discussion which will take advantage of research and development and are presumed to be available on the 2026 timescale. However, to our knowledge, few ideas for the module of opportunity consider expanding or improving the non-neutrino-beam physics program. 

Dual phase LArTPCs have also been successfully deployed as Weakly Interacting Massive Particle (WIMP) dark matter detectors, though currently of a size significantly smaller than the dedicated neutrino experiments. The DarkSide-50 experiment~\cite{agnes2018darkside} used $\sim 50$~kg of low radioactivity argon to set a background-free limit on dark matter. The DarkSide-20k~\cite{DarkSide-20k} and ARGO~\cite{DarkSide-20k}  experiments plan to continue this program down to the so-called neutrino floor. Liquid xenon TPCs~\cite{xenon1t} have also provided the world's most sensitive searches for WIMP dark matter at high masses.

This paper proposes that addressing issues of light collection and radioactive backgrounds with sufficient seriousness provides a viable path to making the DUNE detector sensitive to the detection of nuclear recoil interactions of high mass WIMP dark matter. We discuss the possibility of a detector filled with 17 kt of low radioactivity underground argon that allows for the usual 10 kt fiducial volume and an inner, densely instrumented, self-shielded 1 kt volume of liquid argon for WIMP dark matter detection. Estimations of the radioactive backgrounds yield the conclusion that this detector can be a competitive dark matter detector on a schedule that expedites the world program at a reasonable cost, while preserving the main physics charge of DUNE.
We note that improvements to the light collection and radioactive backgrounds could also improve the sensitivity to supernova and solar neutrinos, further expanding the physics scope of the experiment.

\section{Detector Requirements for WIMP Dark Matter in a DUNE-like detector}
We require the ability to achieve a $\sim$ 100 keV nuclear recoil (keV$_r$) threshold and net backgrounds of $\cal{O}$(10) events or lower from gammas and neutrons over a 3 kt$\cdot$year exposure in order to be sensitive to dark matter in our fiducialized argon. The energy threshold comes from the consideration of what is required to cover interesting parameter space for the WIMP search, though we also explore the increase in sensitivity achievable with lower thresholds and lower backgrounds.

\subsection {The LArTPC Far Detectors envisaged for DUNE}

The proposed DUNE SP far detector is comprised of modular blocks called Anode Plane Assemblies (APAs). One 10 kton cryostat is proposed to consist of 150 APAs. Each APA is comprised of one or two TPCs, which share a common high voltage cathode and two sets of three wire planes. A stainless steel cathode sits in the middle of any two APAs to provide the 500 V/cm electric field for each TPC. Light collection is mounted inside the wire anode planes. Due to the placement of material within the liquid argon we judge the radiopurity requirements for a SP detector to be sensitive to dark matter will be too strict to be achievable on a reasonable budget, so focus this paper on the DP design. 

The proposed DUNE DP far detector is one large (approximately 12x12x60 m$^3$) volume of liquid argon, without APAs or cathodes in the bulk liquid. The drift direction is upward, across 12 meters of liquid argon. The design electric field is the same as for the SP detector (500 V/cm). When ionization electrons arrive at the top liquid surface they are extracted into the gas with a high-field grid that produces an avalanche process, where at least a multiplication of ten is achieved. Charge is collected onto positively biased pads, which are then read out to produce full x,y,z information. 720 PMTs sit on the detector floor in the x-z plane and provide the needed initial time for an event.

\subsection{Proposed DUNE Dual Phase Design Modifications }
\begin{figure}[t]
\begin{centering}
\includegraphics[width=1.00\columnwidth]{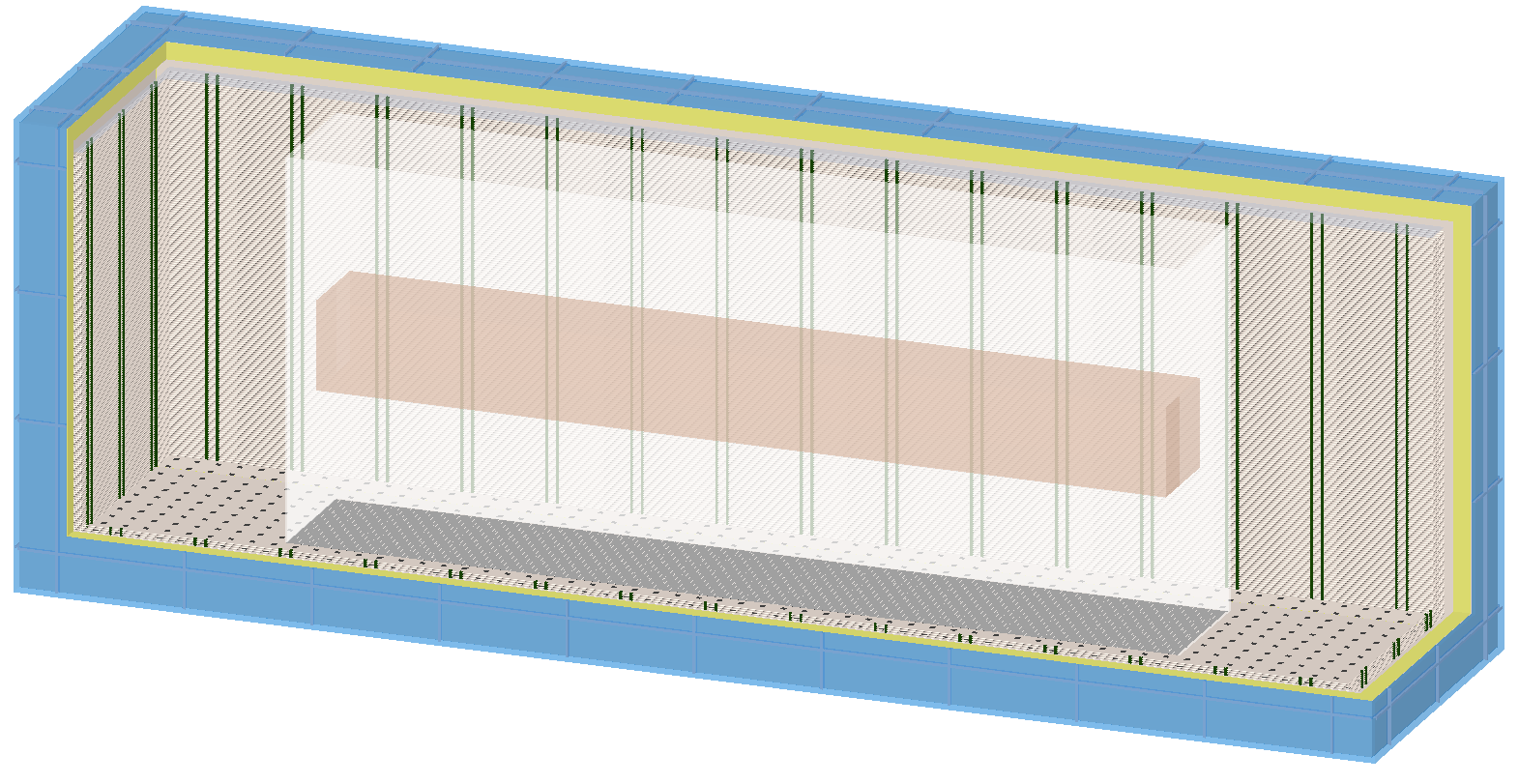}
\par\end{centering}
\caption{Drawing of proposed DUNE-like DP detector for WIMP dark matter searches. The fiducial volume, shown by the beige innermost box, measures 4.2 by 4.2 by 40 m$^3$, with a mass of 1 kt. It is not physically distinct from the bulk argon in which it sits. The tall white box that surrounds the fiducial volume (extended by a meter on the four sides) is formed by 5-cm thick acrylic walls covered in wavelength-shifting foils. The floor of the detector within the acrylic box has 75\% photo-coverage with 24x24 cm$^2$ SiPM tiles (black). The floor outside that area has the same coverage as the original proposed DUNE DP detector: a photodetector at each square meter, except we propose using these same SiPM tiles rather than PMTs. Photo-coverage of the ceiling is not shown, but is similar to the floor except that perhaps an ARIADNE-like camera will serve rather than SiPMs in the gas multiplication region. The field cage is unchanged from the DP design, showing the standard resistor divider chains and supports (green) just inside the cryostat. The cryostat itself consists of a 1.2 cm stainless steel inner "cold cryoskin", 1 cm of wood, 76 cm of polyethelyne foam and an outer stainless steel cryoskin. Outside the cryostat a coarse structure of I-beams and 40 cm of water (blue) are shown. \label{fig:dunecad}}
\end{figure}

For dark matter detection purposes, we propose using the DP design with its uninterrupted volume of liquid argon, relatively free of radioactive material in the bulk of the detector. The modifications we propose to make are: a highly fiducialized low radioactivity argon target, additional shielding, and improved photodetection systems. We show in Figure~\ref{fig:dunecad} some important elements of our detector design.

We propose a central fiducial volume measuring 4.2 by 4.2 by 40 m$^3$, with mass 1 kt, for WIMP dark matter detection. This shape, with at least 3.5 m of liquid argon on all sides, allows self-shielding from external backgrounds. We will replace the atmospheric argon target in this module with low radioactivity argon as discussed in Section~\ref{argon39}. We add acrylic walls inside the detector, of thickness 5~cm, to act as mechanical support for wavelength shifting and reflective coatings and as an additional shield for neutron events from the cryostat. The acrylic walls are situated 1~m outside of the fiducial volume to allow for improved neutron multi-site scattering rejection and the inner volume is open in the vertical direction to allow electron drift.

Light collection in the inner acrylic volume of the detector is enhanced in two ways. We plan to cover the inner walls of the acrylic with wavelength-shifting reflective foils, retaining the scintillation light within this volume. We also propose replacing the photomultiplier tubes with Silicon PhotoMultiplier (SiPMs) and increasing the effective area covered from 3\% to 79\% to improve coverage of the initial scintillation light (S1) (see Section~\ref{photoncount}), which can be used to provide pulse shape discrimination of electron versus nuclear recoils. 24x24 cm$^2$ tile SiPMs as developed for DarkSide-20k would provide a suitable technology for this~\cite{darkside_sipm}. The motivation for the size of the densely-instrumented inner volume (independent of the size of the fiducial volume) is largely cost-based; increasing the size of the inner volume and adding additional photodetection would increase the sensitivity to low energy interactions.

In order to achieve the low-energy thresholds required for sensitivity to nuclear recoils from dark matter interactions we require the dual phase mechanism of charge extraction from liquid into gas and amplification in the gas region.  However, we suggest to measure the ionization charge through the read out of electroluminescence photons (S2), produced by accelerating electrons through the gas phase. Such a strategy enables lower energy thresholds and simpler instrumentation than for charge readout, and we simultaneously improve the overall light collection for the scintillation light (S1). Thus, for the S2 region we remove the charge readout pads in favor of either SiPMs or ARIADNE cameras~\cite{ARIADNE}. We point out that the latter camera is being developed in an intensive R\&D program with the express purpose of competing to serve as the DUNE DP S2 light readout, rather than charge. ARIADNE appears to be able to easily provide the necessary spatial resolution for  multi-site nuclear recoil detection and at a fraction the cost needed for the same SiPM coverage. Finally, external water shielding of 40 cm (as discussed in Reference~\cite{beacom_bkd}) allows reduction of neutrons from the external rock.

\subsection{Backgrounds}
In this section we discuss the radioactive backgrounds and reduction strategies that must be addressed to construct a viable WIMP dark matter experiment in a DUNE-like module.

\subsubsection{Neutrons}
Neutrons are a dangerous background for direct dark matter detectors, because they can induce nuclear recoils that mimic the expected signal from WIMP-like interactions. Neutrons are also a source of background for any nascent DUNE solar and planned supernova neutrino program, as neutron captures on argon produce a 6-9 MeV $\gamma$ that mimics $\nu_e$ charged current reactions. Thus, while neutron amelioration benefits will accrue to any DUNE low-energy neutrino program, we concern ourselves in this paper with reducing and removing neutron-induced nuclear recoils. 

The main sources of neutron backgrounds are cosmogenic and radiogenic. Interactions of cosmic ray muons in the surrounding rock and detector materials can produce high energy neutrons that are difficult to shield. The cosmogenic neutron background is considered in~\cite{beacom_bkd} and may be addressed effectively by tagging the muon and/or fiducializing, as we are doing aggressively. 

Neutrons can also be produced through spontaneous fission or \alphan ~interactions, where the $\alpha$ particle is typically produced in the decay chain of uranium and thorium present in surrounding materials.  For this potential DUNE module we propose to add 40 cm of water outside the cryostat lining, informed by other studies that indicate solar neutrino measurements will require this neutron shielding~\cite{beacom_snu}, as well as our own studies detailed in Section~\ref{sec:nba}. This simple tactic reduces external neutron penetration from the outer rock effectively. In Section~\ref{sec:nba} we also consider the dominant internal neutron contribution from the stainless steel forming the inner cold cryoskin.  

\subsubsection{Betas from \artn}~\label{argon39}
Argon extracted from the atmosphere contains the radioactive isotope \artn~at a specific activity of roughly \SI{1}{\becquerel\per\kgar}. \artn~is a pure $\beta$ emitter with an endpoint energy of \SI{565}{keV}, and therefore, for large liquid argon experiments such as DUNE, the low energy spectrum is dominated by \artn~decay. There are two mitigation strategies that we propose to use to achieve the low background levels necessary for dark matter sensitivity: the use of argon derived from underground sources and pulse shape discrimination.

%\paragraph{Underground Argon}
Since \artn~is primarily produced by cosmogenic activity, argon derived from deep underground (underground argon or \uar) can have extremely low levels of \artn. \uar~derived from gas wells in Colorado has been employed in DarkSide-50~\cite{DarkSide-50}, and was demonstrated to have an $^{39}$Ar activity roughly 1400 times lower (\SI{7.3E-4}{\becquerel\per\kgar}) than the activity in atmospheric argon. DarkSide-20k plans to use this same source of argon. It is clear a new source or \uar~is required for any tens-of-kt liquid argon experiment. A major U.S. gas producer/supplier estimates~\cite{privateconv:gaspro} that approximately 5 kt/year of \uar~may be deliverable at a cost as low as three times that of atmospheric argon. An expression of interest from a collaboration is required for samples to be delivered and the next tests to be conducted: that is, confirming the low activity of $^{39}$Ar. We note that if the proposed central acrylic box can be made leak-tight, \uar~can be deployed only within the dark matter target volume, reducing the amount of \uar~required from 17 kton to 4.3 ktons. An additional benefit to the DUNE low-energy neutrino program of employing underground argon, apart from dark matter physics, is that the reduction of the $^{39}$Ar background activity allows for far superior matching of detected light and charge activity (particularly for the sub-10 MeV range) than is currently achievable and the lower background and trigger rate could also  be beneficial to supernova detection \cite{scholberg2018low}.

%\paragraph{Pulse Shape Discrimination (PSD)} 
Pulse shape discrimination (PSD) from the time profile of argon scintillation light can be used to discriminate nuclear recoil interactions (as expected from WIMP dark matter interactions) from electron recoils with extremely high efficiency. A simple discrimination parameter that is often used is \fnt, which is the fraction of scintillation light that is detected in the first ninety nanoseconds compared to the total amount of scintillation light. For the applied electric fields and energy range of interest to a large LArTPC like DUNE, the average value of \fnt~for electron recoils is roughly 0.3, while for nuclear recoils it is roughly 0.7. The width of the \fnt~distributions, which determines the discrimination power, is strongly dependent on the light collection efficiency of the detector, as the more scintillation light is detected, the more accurately \fnt~can be determined. The DarkSide-50 detector had a light yield of 7.9 photoelectrons/keV$_{ee}$ at null field, and at its operating drift field of \SI{200}{\volt\per\cm} achieved a discrimination factor of $> 1 \times 10^7$, with more than 50\% nuclear recoil acceptance above $\sim 100$ detected photons ($\sim$ 47 keV$_r$) \cite{DarkSide-50}. Through simulations detailed in Section~\ref{photoncount} we show that sufficient light collection for PSD is achievable with an increase in photon detection coverage compared to the current DUNE design. We note that additional discrimination between nuclear and electron recoils can be obtained from the ratio of the ionization to scintillation signal (S2/S1) \cite{benetti2008first}, but it has not been applied here.

There is one further problem that must be overcome: pileup of accidental \artn~decays between the S1 and S2 signals of an individual event, which will make it difficult to associate the correct S1 and S2 signals with each other. The maximum drift time in the current design of the DUNE DP module is roughly 7.5 ms at a drift field of 500 V/cm and (for the assumed \artn~activity of \SI{7.3E-4}{\becquerel\per\kgar} in the underground argon) the rate within the 4.3 kton acrylic box is expected to be 3.1 kBq. The average number of pileup events for interactions at the center of the fiducial volume (drift time $\sim$ 3.8 ms) is therefore 12 events. This pileup can be reduced by using the spatial distribution of S1 scintillation light on the top and bottom photosensors to match it to the observed relative timing and x-y location of S2 signals. If such pattern matching algorithms do not work efficiently enough, one could further optically divide the acrylic box with intermediate reflective foil separators (with little effect on the backgrounds).

\subsubsection{Gammas}
Gamma rays produced by decays of radioactive isotopes in detector materials are another source of backgrounds. The chosen fiducial volume has at least 3.5 m of liquid argon surrounding it on each side which serves to greatly attenuate the gamma ray backgrounds from outside the argon. As higher energy gammas are more penetrating than those at lower energy, we will consider the most abundant high energy gammas from both $^{40}$K and $^{208}$Tl decays.  As shown in Section~\ref{sec:gamma_bkgd_calc}, the residual gamma ray background is negligibly small compared to the internal \artn~background and can be easily removed through PSD. Note that the light-tight acrylic box prevents scintillation light emitted by external gamma ray interactions from affecting the reconstruction of events occurring within the internal volume.

\subsubsection{Radon}
Radon ($^{222}$Rn) is a dangerous background for nearly all low background detectors due to its high mobility as a chemically inert gas, its relatively long half-life (3.8 days), and its broad spectrum of radioactive decay products and energies. Radon emanated from detector materials in direct contact with the argon can enter the active volume where the decays of its daughters can contribute to the background. The contributions from the dominant source of backgrounds are considered in Section~\ref{sec:radon_calc}. We do not consider direct $\alpha$ backgrounds in this study, relying on the difference in energy and scintillation time profile to remove these events. It is possible that geometrical features near the inner detector surfaces may cause effects that shift these events into the region of interest, however position reconstruction and fiducialization will remove this contribution.

The use of an inline cryogenic radon trap in the argon recirculation system can greatly reduce any radon emanated by the warm sections of the recirculation system, as has been demonstrated by the DarkSide-50 experiment \cite{agnes2015first} which achieved a bulk radon concentration of less than 2 $\mu$Bq/kg \cite{DarkSide-20k}. We use this as our baseline assumption for the radon concentration in the proposed detector, though we note that the smaller surface-to-volume ratio of larger detectors could lead to a significantly lower radon concentration level. For example, the DEAP-3600 detector achieved a radon concentration level of 0.15 $\mu$Bq/kg \cite{DEAP-PRD}.

%The radon level in a detector the size of DUNE is expected to be much smaller due to the much larger volume-to-surface ratio.

\subsubsection{Neutrinos}
Neutrinos can be a significant source of backgrounds for large dark matter detectors. For liquid-argon based detectors, neutrino-electron scatters can be easily identified and rejected using PSD. The most dangerous source of background is therefore coherent scattering of neutrinos with the argon nuclei, which are indistinguishable from WIMP-induced nuclear recoils and are therefore an irreducible background. In the energy range of interest for this proposed detector, only atmospheric neutrinos and the diffuse supernova neutrino background are energetic enough to produce nuclear recoils above threshold. The rates of these events are estimated in Section~\ref{sec:neutrino_calc}.

\section{Calculations}

In this section we describe the calculations made to explore the viability of a DUNE-like module as a dark matter detector.

\subsection{Neutron Background Amelioration}~\label{sec:nba}
We propose to ameliorate the neutron background in the following ways: we employ 40~cm of external water tanks outside the cryostat to reduce neutrons from the external rock, we add 5~cm of internal acrylic shielding around the inner volume, we utilize the self-shielding of liquid argon and fiducialize down to an inner kt of argon, and we employ a reconstruction technique that allows to detect and remove multi-site neutron-induced nuclear recoil events. 

%%\rs{\begin{itemize}
%%   \item Describe Geant4 setup (geometry, active volume definition)
%%   \item Neutron spectrum, starting point, details. 
%%   \item Implementation of multi-site rejection, energy threshold
%%   \item Results
%%\end{itemize}}

\begin{figure}[ht]
\begin{centering}
\includegraphics[width=0.90\columnwidth]{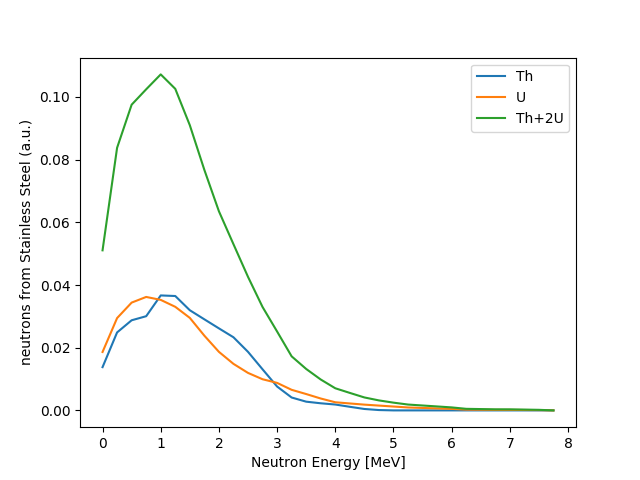}
\end{centering}
\caption{Neutron spectra for  U and Th decays chains in stainless steel from~\cite{vk}, used in this study (normalized in the ratio 2:1) for cold cryoskin neutron emanation.
\label{fig:neutrons}}
\end{figure}

For this study we set up a simple Geant4 10.05~\cite{Geant4} geometry, consisting primarily of a large 17 kt rectangular prism of liquid argon surrounded by a cryostat consisting of a 1.2 cm stainless steel inner "cold cryoskin", 1 cm of wood, 76 cm of polyethelyne foam, and a further outer 1.2 cm stainless steel cryoskin. The full geometry details are shown in Figure~\ref{fig:dunecad} and described in its caption.  A standard physics list is employed, with neutron elastic scatters obeying {\tt G4HadronElasticPhysicsHP}. Neutrons were launched from external planes representing rock neutrons as well as from the 1.2 cm stainless steel of the inner cryostat. The neutron energy spectra (see for example Figure~\ref{fig:neutrons}), were taken from~\cite{vk} and normalized in a 2:1 ratio of \ur:\thr. The neutron activity from the rock is assumed to be $1.0 \times 10^{-5}$~n/cm$^{2}$/s~\cite{cavernRock}. We use a relatively high total activity of $2.0 \times 10^{-9}$~n/cm$^3$/s for the overall normalization in steel, based on the assumption that radiopurity in a large LArTPC is unlikely to reach levels of dedicated dark matter experiments. This compares to the much lower $4.2~(2.7) \times 10^{-11}$~n/cm$^3$/s from  concentrations of 2.8 (1.4) mBq/kg of \ur~and 0.8 (0.82) mBq/kg of \thr~ for DarkSide-20k~\cite{agnes2018darkside} (LZ~\cite{LZ}).

\begin{figure}[ht]
\begin{centering}
\includegraphics[width=0.90\columnwidth]{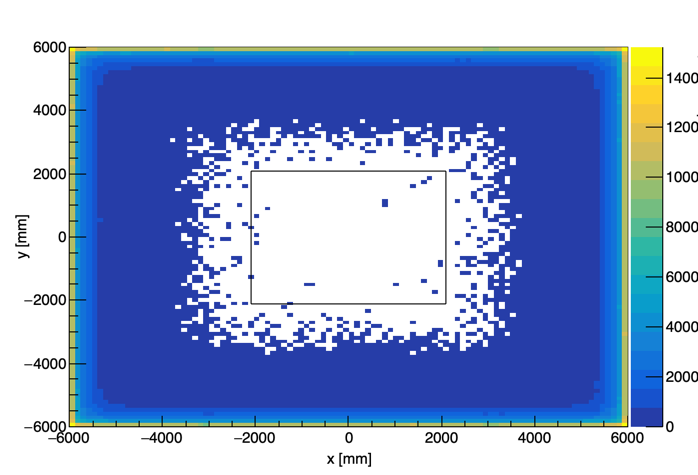}
\end{centering}

\caption{The x-y (transverse to beam) location for a neutron interaction point in which a nuclear recoil above 100 keV occurs and for which the neutron originated from the cold cryoskin. The fiducial volume is superimposed. The z extent (beam-direction) is constrained to the fiducial 40m. This is a 15 kt-year simulation, using a stainless steel activity as described in text. }
\label{fig:neutrons_xy}
\end{figure}

Nuclear recoils above a prescribed energy threshold, usually 100 keV, were selected. Multi-site rejection was identified as a strategy to separate true single-site WIMP scatters from multiple neutron-induced interactions. A neutron scattering more than once in the detector will produce a single S1 scintillation signal (within the time resolution of the photosensors) but can produce more than one S2 ionization signal depending on the relative separation of the interaction points. A dual phase TPC, such as the DUNE DP detector, allows excellent position resolution. Transverse and longitudinal diffusion in the liquid argon across drifts of $\sim10$~m are expected to be far less than 1~cm. The time-sampling of the arriving drift electrons is easily high enough to preserve the longitudinal spread, whereas a few cm transverse resolution from the S2 light detectors -- easily met by the ARIADNE system and also by a SiPM solution at higher cost -- allows sufficient resolution to reconstruct distinct nuclear scatters. If a second above-threshold unpaired S2 signal is detected in the region within the acrylic box, we remove that event from consideration as a neutron multi-scatter cut. We presume we have a 20~mm resolution in the dimension perpendicular to the drift for this two-site tagging. We calculate the resulting neutron background which survives propagation and multi-site rejection in the inner 1 kt. Figure~\ref{fig:neutrons_xy} shows a 15 kt-year simulation of the sites of a nuclear recoil of at least a 100 keV, demonstrating the self-shielding efficacy of the argon itself.

Through both simulation and the study of~\cite{beacom_bkd} we calculate that 40 cm of external water moderator and a 100 (75) keV nuclear recoil threshold removes essentially all external rock neutrons, giving a residual rate of 0.1 (1.6) counts/3 kt-yr. From simulations we also learn that for a 100 keV threshold less than one event from the stainless steel inner cold cryoskin survives into our inner kiloton of liquid argon per year of exposure. A 75 keV nuclear recoil threshold admits 9.55 events. We find the multi-site rejection gives a factor of 30\% (50\%) rejection at 100 (75) keV threshold. Multiplying by 0.5 for nuclear recoil acceptance and by 3 for a 3 kt-yr exposure we arrive at 1.02 (14.2) events for a 100 (75) keV threshold. Achieving these backgrounds imposes a not-unreasonable radiopurity constraint on the cryostat steel. It should be noted if the radiopurity of stainless steel from dark matter experiments could be achieved in a LArTPC like the one proposed here, a further factor 100 reduction in neutron production would lead to the removal of this background.

\subsection{Photon Counting}~\label{photoncount}
Light detection in the currently designed DUNE dual phase detector is relatively limited and is dedicated to timing measurements (T$_0$) of 
%non-beam 
events in the drift direction. The system includes 720 8" Hamamatsu R5912-20Mod PMTs, distributed uniformly across the bottom surface of the detector, resulting in a photocathode coverage of about 3\% on this surface. The target dual phase photon detection light yield from ProtoDUNE data and simulation is an average of 2.5 photoelectons (pe)/ MeV~\cite{TDR}. To take advantage of the pulse shape discrimination information in the initial scintillation (S1) light, this light yield is insufficient and must be increased to closer to 1000 pe/MeV.

To explore the effects of increasing the photon detection system coverage on the light yield a pseudo-Monte Carlo model of photon propagation in the detector was created. Argon scintillation events were created randomly throughout the detector fiducial volume with random momentum vectors, assuming prompt scintillation of 1250 photons per 100 keV, as measured by the SCENE collaboration for fields of 500 V/cm~\cite{cao2015measurement, cao2014study}. Individual photons were extrapolated to the edges of the detector, where photons were either absorbed and lost at the wall, or triggered the photon detector assuming a 50\% wavelength shifting probability (from geometric acceptance arguments) and using measured quantum efficiencies (QE) of the PMTs of 14\%. Attenuation was assumed to be negligible in this study. Measurements of attenuation length are highly dependent on argon purity, and attenuation lengths of 30-60 m have been achieved with parts-per-million levels of nitrogen and kilometer-scale attenuation lengths with parts-per-billion (ppb) levels~\cite{arAtten1}. Since the presence of impurities can strongly suppress the scintillation light from the long-lived triplet state required for pulse shape discrimination, ppb levels of impurities will be required for this module. This corresponds to less than 0.5\% reduction in the collected light. Using a Rayleigh scattering length of 55cm, and assuming a 50\% probability that a photon scatters backwards away from the photosensors during such a scatter, the DUNE DP simulation result of 2.5 pe/MeV was reproduced to within 20\% accuracy.

A study of options to increase the photon detector system coverage to reach light yields of 1 pe/keV, was made. To reach this target several improvements must be made:
\begin{itemize}
    \item Increase reflectivity of internal surfaces of the detector. A foil of 97\% reflectance was assumed at the edge of the inner volume attached to the inner wall of the acrylic shield. This was simulated as a specular reflector and photons were allowed to reflect up to 10 times before absorption at the wall or detection at a photosensor. A simplified ray tracing, without Rayleigh scattering, was used for this study.
    \item Enhancement of light coverage. An increase in the geometric coverage at the bottom of the detector was simulated by increasing the packing of the photosensors, and hence the global photocathode coverage. The possibility of using additional light collection at the top of the detector (by replacing the charge readout with SiPMs or the ARIADNE cameras) was also considered.
    \item Increase photon detection quantum efficiency. It was found that even with full geometric coverage on the bottom with the current planned DUNE DP PMTs, the light yield could not reach 1 pe/keV due to the 14\% quantum efficiency of the detector. It is proposed to use SiPMs with 45\% QE as planned for DarkSide-20k.
\end{itemize}

\begin{figure}[ht]
\begin{centering}
\includegraphics[width=0.90\columnwidth]{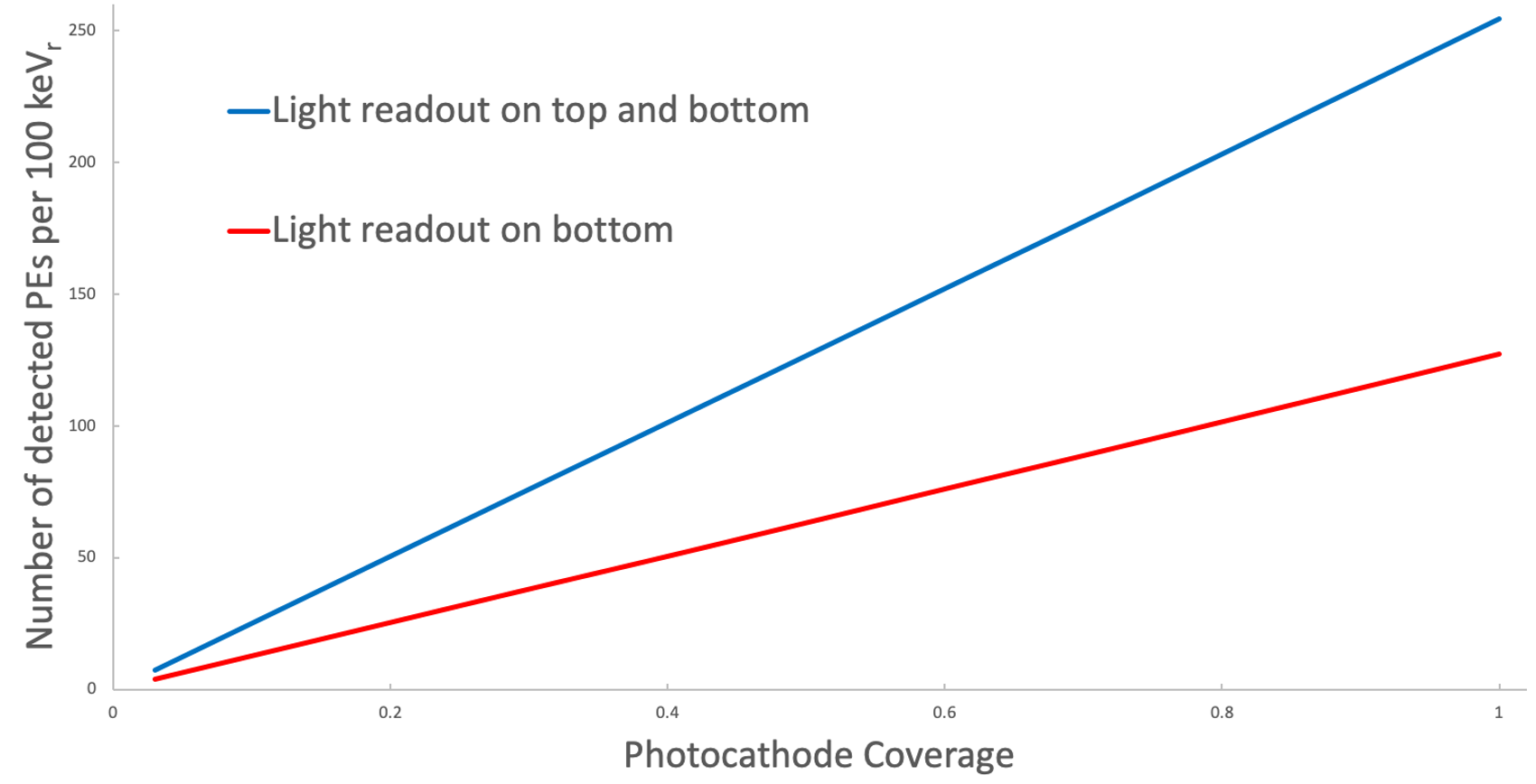}
\par\end{centering}
\caption{Number of observed photoelectrons (pe) per 100 keV$_r$ as a function of geometric coverage of the inner volume. It is assumed readout is via SiPM tiles and that the walls of the inner volume are covered with reflector as discussed in the text. \label{fig:pecoverage}}
\end{figure}

The results of the toy simulation where the refective foils and SiPMs are added are shown in Figure~\ref{fig:pecoverage} as a function of geometric photocathode coverage. It was found that to reach the 1 pe/keV target light yield, photocathode coverage must be $\sim$79\% or $\sim$39\% for the assumptions that light can be read out on the bottom only or top and bottom respectively. This corresponds to a total of $\sim$3600 24x24 cm$^{2}$ DarkSide-20k SiPM tiles, which is within the reach of planned production techniques.

\subsection{Pulse Shape Discrimination}\label{sec:psd_calc}

\begin{figure}[ht]
\begin{centering}
\includegraphics[width=0.90\columnwidth]{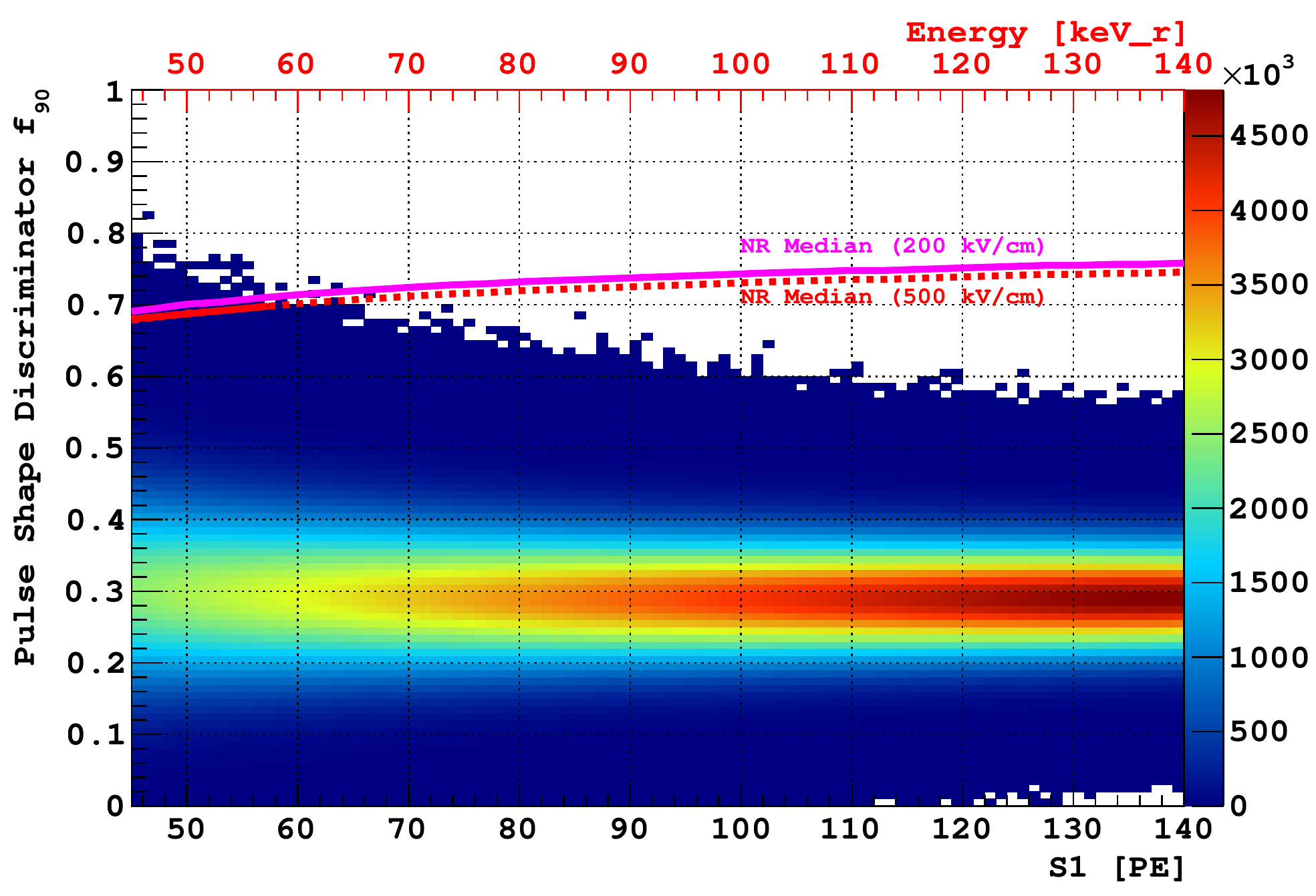}
\par\end{centering}
\caption{Pulse shape discrimination F90 variable as a function of scintillation light S1. The figure shows the \artn~electron recoil band simulated by the Monte Carlo simulation code described in the text. The median of the nuclear recoil band at 200 V/cm and 500 V/cm (extrapolated above 57 keV$_r$) is shown for comparison.  \label{fig:fnt_sim}}
\end{figure}

In order to evaluate the electron recoil rejection power achievable through pulse shape discrimination (\psd) a Monte-Carlo simulation code was used to generate the \fnt~distribution of \artn~events as a function of energy. The statistical simulation starts by generating Poisson-distributed photons at prompt (first 90 ns) and late times (after 90 ns) based on the energy of the interaction, the energy and field-dependent scintillation yields, and \fnt~(prompt/total) medians. The nuclear recoil scintillation yield quenching and \fnt~medians at 500 V/cm were obtained from the SCENE experiment \cite{cao2015measurement, cao2014study} for energies below 57 keV. Above those energies we used a constant extrapolation for the scintillation yield quenching and for the \fnt~medians followed the same energy dependence measured by DarkSide-50 at 200 kV/cm \cite{agnes2018darkside} (as shown in Figure~\ref{fig:fnt_sim}). The electron recoil scintillation yield quenching was taken from the $^{83}$Kr measurements at 500 V/cm by SCENE \cite{cao2015measurement, cao2014study} while the \fnt~medians were obtained from the DarkSide-50 data \cite{agnes2015first} measured at 200 V/cm, due to lack of data at 500 V/cm. The overall light collection efficiency plays a critical role in the discrimination power and was set to 3.75 pe/keV$_{ee}$ at null field to match the 1 pe/keV$_{r}$ (at 500 V/cm field, above 57 keV$_{r}$) light collection estimated in the optical simulations above. The simulation takes into account variations in the recorded number of photoelectrons due to the response of traditional PMTs to single photoelectrons ($\sigma \sim 0.4$ photoelectrons) and digitization noise. We note that the use of silicon photomultipliers instead of traditional PMTs would greatly reduce this contribution. The simulated prompt and late signals are then combined to form the S1 scintillation and \fnt~variables. 

As discussed above, even with the use of underground argon, we expect that \artn~will be the dominant electron recoil background in this detector and have only simulated events with energies drawn from the \artn~beta decay spectrum. The output of this simulation shows good agreement with the measured F90-S1 distributions, dominated by \artn, in the initial DarkSide-50 data using atmospheric argon (see Figure 4 in \cite{agnes2015first}). We note that the simulation does not take into account photon time-of-flight delays that might be relevant in detectors of the proposed size (though it can be mitigated with the addition of optical separators), or events with unusual topologies, such as interactions in detector materials that can produce fast Cherenkov light \cite{agnes2018darkside, koh2018dark}. The most significant potential source of Cherenkov light, from the acrylic walls used to mount the reflectors, is managed in our design by ensuring the reflective foils are opaque and mounted on the inner surface.

We simulated a total exposure of 3 kton$\cdot$yrs, assuming an \artn~rate of \SI{7.3E-4}{\becquerel\per\kgar} as achieved in low-radioactivity underground argon, as shown in Figure~\ref{fig:fnt_sim}. It can be seen that in this simulation no events were produced with \fnt~above the nuclear recoil median for energies greater than 67 keV$_r$. To account for statistical fluctuations as well as some of the simplifications in the model mentioned above, we have assumed that one \artn~event will leak above the nuclear recoil median for a threshold of 75 PE (75 keV$_r$) and 0 events above 100 PEs (corresponding to 100 keV$_r$, or roughly 32 keV$_{ee}$). For reference, we expect roughly 1.6 (1.7) $\times10^{10}$ \artn~decays in this exposure to fall in the 100 (75) - 400 pe range. Since this number is orders of magnitude larger than the contributions from other sources of electron recoils in the fiducial volume of interest, pulse shape discrimination should eliminate all other electron recoil backgrounds.

\subsection{External Gamma Background Amelioration}
\label{sec:gamma_bkgd_calc}
It is expected that gammas from detector materials will be significantly reduced by the planned 1~kt fiducialization. To evaluate potential leakage events into the fiducial volume we used the Geant4 simulation described in Section~\ref{sec:nba} to evaluate this risk. The two most dangerous sources of gamma background, due to proximity to this volume, are likely to be the proposed acrylic box surrounding the fiducial volume, and the photosensors and electronics at the top and bottom of the detector, with the top more susceptible due to the low density gas multiplication region. In order to estimate the potential background contribution we simulated $^{40}$K and $^{208}$Tl, which produce high energy 1.4 and 2.6 MeV gammas respectively, from each of these regions.

We find that for acrylic with a few parts-per-trillion $^{232}$Th content (as achieved by SNO~\cite{acrylic} and DEAP-3600 \cite{deapacrylic}), a 5-cm thick box bordering our inner fiducial volume contributes about 
85,000 events for 3 kt-years in our fiducial volume, almost flat across our choice of thresholds, before applying any pulse shape discrimination.

For gammas emitted from the top of the detector we 
explore $^{40}$K and $^{208}$Tl decays, respectively, 
in our fiducial volume. Given reasonable material radioactivities of 1 Bq/kg $^{40}$K and 10 mBq/kg of $^{232}$Th and a mass density of 0.5 gm/cm$^2$ (e.g. 3-mm thick G-10) we place upper limits of 30 and 4.3 
events/3 kt-years from these two sources, respectively, in the inner 1-kt fiducial volume -- again before application of PSD. For $^{232}$Th decays this follows from non-observation of a single event in the fiducial volume at any of our thresholds for a reasonable simulation job length (a 30$^{\text{th}}$ of 3 kt-years). For $^{40}$K, where our simulations also give zero events,   we scale from the $^{232}$Th result with a ratio of the above activities multiplied by the ratios of attenuation through 3.6m of argon for the two different gamma energies.

%% >>> 30 * 1.0/0.010 * np.exp(-5.0E-2*1.4*3.6*100)/np.exp(-3.7E-2*1.4*3.6*100)
%% 0.0014272582254365829
%%

In conclusion, the rates of electron recoils from both of these sources are negligible compared to the \artn~rate and such events are completely removed through the pulse shape discrimination described in Section~\ref{sec:psd_calc}.

\subsection{Radon Daughter Backgrounds}
\label{sec:radon_calc}

In this section we describe calculations of radon-induced background levels.

\subsubsection{$^{214}$Pb}
$^{214}$Pb is a low energy beta-emitting daugther of radon, whose decays produce electron recoils that fall within the energy range of interest. From a radon concentration at the DarkSide-50 upper limit of 2 $\mu$Bq/kg, 63000 events/(ton$\cdot$yr) from $^{214}$Pb decays would be expected, giving 1.9$\times10^8$ events in the total 3 kton$\cdot$yr exposure. Only a fraction of these events would fall in the energy range of interest, and as shown in Section~\ref{sec:psd_calc}, this contribution is $<1\%$ of the \artn~background and can be easily removed through the application of pulse shape discrimination.

\subsubsection{$^{40}$Ar($\alpha$, n)}
A more dangerous background for large argon-based dark matter detectors, that to our knowledge has not been previously discussed in the literature, is neutrons produced by $^{40}$Ar \alphan~interactions from $\alpha$-emitting radon daughters. The ($\alpha,n$) simulation code NeucBOT \cite{westerdale2017radiogenic} estimates a total neutron yield of \num{3.62E-7}, \num{1.10E-6}, and \num{1.24E-5} n/decay for the alpha emissions of $^{222}$Rn, $^{218}$Po, and $^{214}$Po respectively. For a radon concentration of 2 $\mu$Bq/kg, this corresponds to a neutron rate of roughly \num{1.5E4} neutrons/17.5 ktons/yr in the whole argon module. Geant4 simulations of neutrons drawn from the expected \alphan~spectrum indicate that roughly 3\% or \num{1.3E3} neutrons/3yr of these neutrons will produce a single-scatter nuclear recoil in the 1-kton fiducial volume, and would be the dominant source of nuclear recoil backgrounds in the energy range of interest.

Since these nuclear recoils would occur in prompt coincidence with the \alphan~interaction that produces the neutron, any energy deposits associated with the \alphan~interaction would be detected in coincidence with the nuclear recoil signal, either pushing it out of the window of interest or, if it is from a gamma, making the detected scintillation time profile more like an electron recoil such that it can be removed through PSD. 
The $^{214}$Po decay, which contributes roughly 90\% of the \alphan~neutrons, occurs in delayed coincidence with the preceding $^{214}$Bi decay, with a half-life of 164 $\mu$s. Since the $^{214}$Bi decay typically deposits more energy than the \artn~$\beta$~decay endpoint, the $^{214}$Po decays can be tagged and removed with high efficiency, though the coincidence time window will have to be set taking into account the corresponding loss of livetime.

The $\alpha$ decays of interest all have energies above 5 MeV. Should the $\alpha$ deposit even a small fraction of its energy in the argon before inducing the \alphan~interaction, the associated scintillation light would push the detected event outside the 100-400 photoelectron window proposed for the dark matter search. 

Even if the $\alpha$ does not deposit significant energy in the argon, there is a high probability that the \alphan~interaction will proceed to an excited state of the $^{43}$Ca nucleus, likely producing a coincident gamma ray. The Q-value for the \alphan~reaction is -2.37 MeV and the lowest excited state of the $^{43}$Ca nucleus is only 373 keV. TALYS 1.95 \cite{koning2007talys} simulations of 5.5 and 6.0 MeV $\alpha$s (corresponding to $^{222}$Rn and $^{218}$Po respectively) indicate that 88\% and 91\% of the \alphan~interactions go to an excited state of the $^{43}$Ca nucleus. 

A quantitative estimate of this background requires a full simulation of the $\alpha$ interaction in the argon to estimate the amount of energy deposited in the argon before the \alphan~interaction, an accurate modeling of the resulting excited states and decay products of the $^{43}$Ca nucleus, the conversion of these energy deposits into scintillation and ionization, and a model of the detector response and tagging efficiency. This is beyond the scope of this current work and we have assumed that all of the neutrons generated by $^{40}$Ar\alphan~can be rejected. However, given the dangerous nature of this background, this should be more thoroughly evaluated for all large scale liquid argon dark matter detectors.

\subsubsection{Radon Plate-out}
We note that decay of radon in the air surrounding detector components can lead to the plate-out of radon daughters on the surface. Alpha emitting daughters can then induce the production of neutrons through \alphan~interactions on the surfaces (e.g. $^{210}$Pb alphas interacting with $^{13}$C in the acrylic). The magnitude of this contribution compared to that from \ur~and \thr~in the bulk of the materials depends on the specifics of the materials used, the location and storage of the detector materials, and handling procedures during assembly, and is not included in these preliminary estimates. 

\subsection{Neutrino Backgrounds}
\label{sec:neutrino_calc}
We calculate an expected background contribution from neutrino nuclear coherent scatters at various thresholds from references~\cite{strigari1} and \cite{strigari2} The standard model cross-section is integrated with the Honda neutrino flux~\cite{honda} above the desired nuclear recoil threshold energy. We arrive at 19.3 (25.7) events for a threshold of 100 (75) keV events for our three kton-year exposure, dominated by the contribution from atmospheric neutrinos. We take half of that, 10 (13) events, as our irreducible background given our defined 50\% nuclear recoil acceptance from pulse shape discrimination. A 20\% error is assigned in these references, though as with all our backgrounds we take the central value.

\section{Sensitivity for Dark Matter Detection}

We use a WIMP sensitivity calculation~\cite{loer-code} that takes into account the kinematic details of WIMP scattering off heavy target nuclei, including nuclear form factors and effects\footnote{We use $v_{\tt escape}=544$ km/sec, $v_{\tt dispersion}=220$ km/sec and $v_{\tt rotation}=244$ km/sec and a dark matter density $\rho=0.3$ GeV/c$^2$/cm$^3$.} of the Earth's motion~\cite{lewinsmith}. For simplicity we assume a constant 50\% nuclear recoil acceptance at all energies, though an energy-dependent acceptance can be used to take advantage of the strong energy-dependence of the PSD. In Figure~\ref{fig:Sensitivity} we show experimental limits from previous argon-based experiments and the current best limit at high masses from XENON1T~\cite{xenon1t}. We also show the projected sensitivity reach of the liquid argon  DarkSide-20k~\cite{DarkSide-20k} experiment, along with that of the ARGO experiment~\cite{DarkSide-20k}, which is the planned follow-on to DarkSide-20k. We present three potential sensitivities from our large DUNE-like detector after running our 1~kt fiducial volume for 3 years. Such a strategy could produce a physics result on the timescale of the DarkSide-20k result.

Lastly, we comment on optimizing the fiducial volume. Upon varying the transverse dimension for several choices of fiducial volume length, the geometry that maximizes signal to background uncertainty at the 100 keV$_r$ threshold was found to be roughly 5.3 m x 5.3 m x 52 m, for a total mass of 2.0 kt. The increase in sensitivity for this volume, with its attendant increase in background, compared to our 1 kt choice is quite small ($\sim$ 20\%). We therefore use a conservative 1 kt fiducial volume of the dimensions described for all studies in this paper.

\begin{table}[t]
\begin{tabular}{ l c c c c c}
%%%%\multicolumn{4}{c}{\bf Background in 10 kt-yr 100 keV threshold} \\
Background  & Amelioration strategy & & \multicolumn{3}{c}{Counts/3 kt-yr} \\
 & & & 100 keV$_r$ & 75 keV$_r$ & 50 keV$_r$*\\
\hline
  neutrons from  & external 40 cm water & & 0.1 & 1.6 & 13  \\
  external rock & self-shielding, multi-site rej. & & & & \\[8pt]
 neutrons from & self-shielding, & & 1.02 & 14.2 & 2 \\
  cold cryoskin steel & acrylic, multi-site rej. & &  & & \\[8pt]
  $^{40}$K gammas & self-shielding, PSD & bPSD: & \multicolumn{3}{c}{$<4.3$} \\
  from detector top & & aPSD: & 0 & 0 & 0 \\[8pt]
$^{208}$Tl gammas & self-shielding, PSD & bPSD: & \multicolumn{3}{c}{$<30$} \\
from detector top & & aPSD: & 0 & 0 & 0\\[8pt]
$^{208}$Tl gammas & PSD & bPSD: & 8.1$\times10^4$ & 8.5$\times10^4$ & 8.9$\times10^4$ \\
from acrylic & & aPSD: & 0 & 0 & 0 \\[8pt]
$^{214}$Pb & PSD & bPSD: & \multicolumn{3}{c}{$< 1.9$$\times10^8$} \\
 from radon & & aPSD: & 0 & 0 & 0 \\[8pt]
$^{40}$Ar\alphan & coincident tagging & & 0 & 0 & 0 \\
from radon& (see Section~\ref{sec:radon_calc}) & & & & \\[8pt]
$^{39}$Ar betas & \uar, PSD & bPSD: & $1.6$$\times10^{10}$ & $1.7$$\times10^{10}$ & $1.8$$\times10^{10}$\\
in argon & & aPSD: & 0 & 1 & 1 \\[8pt]
atmospheric & none & & 10 & 13 & 17 \\
neutrinos & & & & &\\
\hline
Total & & & 11 & 30 & 33\\
\hline
\end{tabular}
\caption{Backgrounds considered and estimated count rates in the inner 1 kt fiducial volume over 3 years for different energy thresholds.  These estimates presume 40 cm water tanks between the rock wall and outer cryostat, the use of underground argon with an \artn~rate of \SI{7.3E-4}{\becquerel\per\kgar}, multi-site rejection with 20 mm resolution, and a 50\% nuclear recoil acceptance. Electron recoil backgrounds are shown before (bPSD) and after (aPSD) application of PSD. *The numbers listed for a 50 keV$_r$ threshold require a change in the detector configuration; see text.}
\label{table:DUNE}
\end{table}

\section{Results and Discussion}
\begin{figure}[t]
\begin{centering}
\includegraphics[width=0.90\columnwidth]{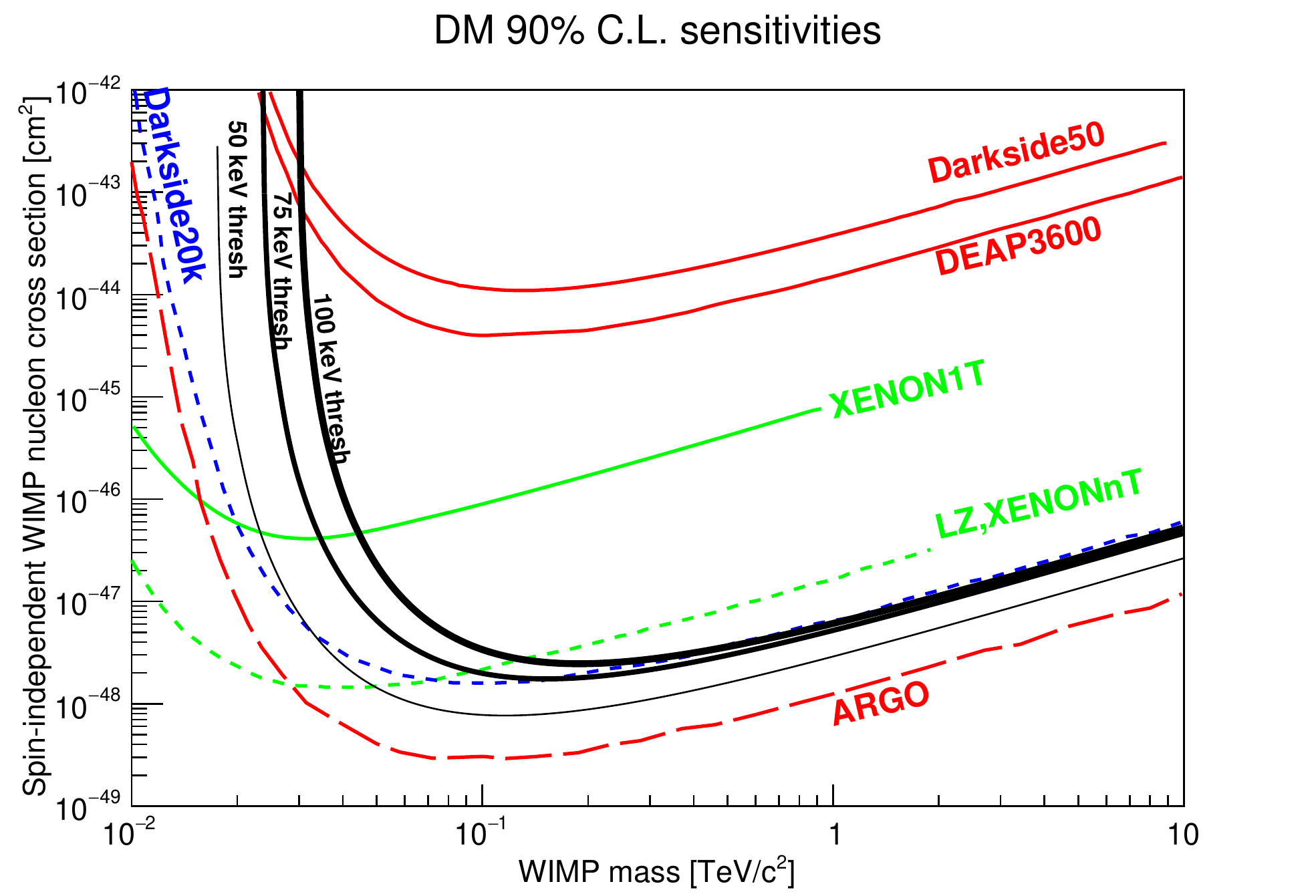}
\par\end{centering}

\caption{Shown are the achieved 90\% c.l. sensitivities in solid colored lines for various experiments for WIMP Dark Matter
%%amaudruz2018first
searches~\cite{DEAP-PRD,DarkSide-50,xenon1t}. Dashed lines for proposed argon~\cite{DarkSide-20k} and xenon~\cite{LZ, darwin} experiments are overlaid. The DarkSide-20k exposure is 20 tons over a planned 10 years for a total of 200 ton-yrs, while our proposal for a DUNE-like module is 3 years of a 1 kt mass for a 3 kt-yr exposure. The ARGO exposure is a planned 300 tons over ten years for a total of 3 kt-yr. Two possibilities for a DUNE-like module 4 for two thresholds and expected backgrounds, as discussed in the text, are shown in black lines. A third, aspirational, DUNE module 4 sensitivity for  a 50 keV threshold  is also drawn as a thinner line.   \label{fig:Sensitivity}}
\end{figure}

 The backgrounds at each threshold are summarized in Table~\ref{table:DUNE} and the sensitivity results for the three thresholds studied are shown in Figure~\ref{fig:Sensitivity}. We judge the background scenarios and thresholds as progressively more ambitious, but  achievable if light collection and radiopurity standards are increased as required. 
 
 For a 100 keV nuclear recoil threshold the gamma backgrounds are small with respect to \artn, and negligible after application of PSD, while the neutron backgrounds may be controlled by a reasonable radiopurity requirement for the cryostat steel (at a less stringent level than leading dark matter experiments), in combination with exterior water tanks. For a 75 keV$_r$ threshold we show that the neutron background is still sub-10 events, and PSD continues to give the necessary rejection factor. We find  atmospheric neutrinos are the dominant and irreducible background for a 100 keV$_r$ threshold and are on equal footing to the cryostat steel neutrons for a 75 keV$_r$ threshold.

Finally, the most ambitious sensitivity curve we show for this study is at a 50 keV$_r$ threshold.  The contribution of neutrons from the cold cryoskin at this threshold is 174 events. If we require the cryostat steel to be a factor 80 cleaner, at levels planned for LZ and DarkSide, we can lower this neutron background to roughly 2 events. At a 50 keV$_r$ threshold the contribution from external rock neutrons only increases to 13 events as a result of the fact that at this low threshold our multi-site tagging removes more than 90\% of events. Further, if we increase the photo-coverage to roughly 60\% of the top and bottom surfaces of the detector we can increase the light collection to 1.5 pe/keV$_r$, which allows for only one electron recoil leakage event after PSD. Assuming that all gamma backgrounds remain sub-dominant to \artn~at this threshold, we obtain a total background contribution (including atmospheric neutrinos) of 33 events, leading to the {\em aspirational} sensitivity curve shown on our plot.

We see from Figure~\ref{fig:Sensitivity} that due to the higher threshold (compared to dedicated dark matter detectors) this experiment is mostly sensitive to WIMP masses larger than 30 GeV. Though a 50 keV$_r$ threshold for a DUNE-like detector module will be very difficult to achieve, given the sheer size of the detector and industrial materials that must be employed, we believe a 100~keV$_r$ (and even a 75~keV$_r$) threshold is achievable with some reasonable radiopurity controls. Further, we have some confidence based on our own toy Monte Carlo and~\cite{privateconv:flavio} that 100 photons at 100 keV$_r$ can be achieved with upgraded dense optical coverage, as we suggest here. The resulting sensitivity curves show an interesting coverage with respect to DarkSide-20k, allowing a confirmation or refutation of a high mass signal between the two experiments. %Further, we show that an ambitious background reduction and an intermediate threshold to the two just mentioned, in fact, allow to get to the neutrino floor in a manner akin to GADMC's plans.

%\todo{It is worth pointing out that there are other opportunities in the DUNE program, as described in the Technical Design Report~\cite{TDR}, that could benefit significantly from improvements to light collection and radioactive backgrounds in both the SP and DP detectors. The supernova explosion detection efficiency is currently estimated to be lower than had been specified in the Conceptual Design Report, particularly at low energies. The proposed module, with its low energy threshold for the inner fiducialized volume, could even serve as the supernova event trigger for the full 40 kt DUNE far detector complex. The solar neutrino program, recently being envisioned for DUNE~\cite{beacom_snu} would benefit from a lower threshold than the current $\sim10$ MeV; formation of light "flashes" for determination of event time (T0) in all manner of non-beam physics events is relatively poor for low energy events, especially when events occur far from the light collectors. These are issues that have little bearing on the main DUNE beam neutrino physics program but which, if addressed properly, might significantly expand the physics scope of the experiment.}

It is worth pointing out that there are other opportunities in the DUNE program, as described in the Technical Design Report~\cite{TDR}, that could benefit significantly from improvements to light collection and radioactive backgrounds in both the SP and DP detectors. The energy threshold for supernova detection in the fiducialized volume could be lower than currently envisioned, and perhaps this volume could even serve as the supernova event trigger for the full 40 kt fiducial DUNE far detector complex. The solar neutrino program, recently suggested for DUNE~\cite{beacom_snu}, would benefit from a lower threshold than the $\sim$10 MeV considered there. Lastly, the process of forming light "flashes" and determining $T_0$ for events occurring far from the light collectors in any large LArTPC will clearly improve with denser photo coverage and potentially expand the physics scope of the experiment.

\section{Conclusion}
We show in this study that a large LArTPC detector filled with low-radioactivity underground argon can be competitive for high mass WIMP detection, given low radioactivity materials and improved light collection for a self-shielded inner 1 kt of liquid argon within a DUNE-like 10 kt detector. We have some confidence from proprietary discussions that affordable, commercial sources of such underground argon could be obtainable on the timescale relevant to the fourth DUNE module.

We show with a toy Monte Carlo that this improved light detection added to a large LArTPC, like the DUNE dual phase design, is feasible with a denser array of photosensors and reflective foils placed in the argon. Pulse shape discrimination can then be employed to effectively remove electron recoil backgrounds within the fiducial volume. We show that neutrons in our inner 1 kt may be ameliorated to small levels with an outer 40 cm of moderating water and cuts to remove multi-site interactions. We also investigated the improvement in sensitivity possible with aggressive radiopurity controls and even higher light collection efficiency.

With a detector module as described here, a broad low-energy physics program would be enabled that would not perturb the main DUNE neutrino program. In addition to dark matter detection described in this paper, the proposed changes could also lead to more efficient measurements of solar neutrinos and low energy supernova neutrino interactions. 

\acknowledgments
The authors wish to express their gratitude for valuable discussions with Henning Back, Jim Fast, Ben Loer, and Shawn Westerdale,  and for the detector design drawing by Gabriel Ortega. This paper is not published on behalf of the DUNE collaboration. This work was funded under the Nuclear-physics, Particle-physics, Astrophysics, and Cosmology (NPAC) Initiative, a Laboratory Directed Research and Development (LDRD) effort at Pacific Northwest National Laboratory (PNNL). PNNL is operated by Battelle Memorial Institute  for  the  U.S.  Department  of  Energy  (DOE) under Contract No. DE-AC05-76RL01830.

% We suggest to always provide author, title and journal data:
% in short all the information that clearly identify a document.

\end{document}